
%
\font\bigbf=cmbx10  scaled\magstep1

\font\ggreatrm=cmr10  scaled\magstep4

%
%

%

%

%

%

%

%

%
\def\etal{{\it et al.\/}}
\def\tento #1 {\cdot 10^{#1}}
\def\Bigroman#1{\uppercase\expandafter{\romannumeral #1}}
\def\Sggreat{{\hbox{\lower 4pt \hbox{\ggreatrm S}}}}
\def\bvec#1 {{\bf #1}}
\def\simlt{\lower.5ex\hbox{$\; \buildrel < \over \sim \;$}}
\def\simgt{\lower.5ex\hbox{$\; \buildrel > \over \sim \;$}}
\def\rahm #1,#2 {#1^{\hbox{\sevenrm h}}~#2^{\hbox{\sevenrm m}}}
\def\decdm #1,#2 {#1^o~#2'}
%
%
\def\today{%
     \ifcase\month\or January\or Febuary\or March\or April\or May\or
     June\or July\or August\or September\or October\or
     November\or December\fi \enspace\number\day, \enspace\number\year.}
\def\clock{\count0=\time \divide\count0 by 60
    \count1=\count0 \multiply\count1 by -60 \advance\count1 by \time
    \number\count0:\ifnum\count1<10{0\number\count1}\else\number\count1\fi}

\footline={\hss {\rm -- \folio\ --} \hss}


\def\draft#1{\headline={ %
{\bf DRAFT:\ \jobname.tex ~~---~~ version \# #1}
{\sevenrm \hss \today ~~~ Hrs.~\clock}}
  }

%
%

%
\newskip\eightptskip
	\eightptskip = 8pt  plus 1pt minus 1pt

\newskip\sixptskip
	\sixptskip = 6pt  plus 1pt minus 1pt

\def\oneskip{\vskip\baselineskip}

\newskip\hskip
	\hskip = \baselineskip
	\divide\hskip by 2
\newskip\tskip
	\tskip = \baselineskip
	\divide\tskip by 3
\def\halfskip{\vskip \hskip}
\newskip\hoskipup
	\hoskipup = \hskip
	\multiply\hoskipup by -1

\newskip\oskipup
	\oskipup = \hoskipup
	\multiply\hoskipup by 2

\newskip \nullskip
\nullskip = 0pt plus 3 pt minus 3 pt
%
\newskip \ssectskip
	 \ssectskip = \baselineskip
\def\sectvskip{\vskip\ssectskip}
\newskip \ssubsectskip
	 \ssubsectskip = \hskip
\def\subsectvskip{\vskip\ssubsectskip}
\newskip \ssubsubsectskip
	 \ssubsubsectskip = \tskip

%

%
\def\ref{\penalty -100 \par \noindent \hangindent \parindent}
\def\sequel{\hbox to 3truecm{\hrulefill}}
\newif\iflongrefs\longrefsfalse
\iflongrefs

\def\aa #1 #2.{{Astronomy \& Astrophysics} #1, #2.}
\def\araa #1 #2.{{Annual Review of Astronomy \& %
Astrophysics} #1, #2.}
\def\aj #1 #2.{{Astronomical Journal} #1, #2.}
\def\apj #1 #2.{{Astrophysical Journal} #1, #2.}
\def\apjl #1 #2.{{Astrophysical Journal (Letters)} #1, L#2.}
\def\apjsupp #1 #2.{{Astrophysical Journal Supplement  %
Series} #1, #2.}
\def\mnras #1 #2.{{Monthly Notices of the Royal astronomical   %
Society} #1, #2.}
\def\qjras #1 #2.{{\it Quarterly Journal of the Royal astronomical   %
Society} { #1}, #2.}
\def\nat #1 #2.{{Nature} #1, #2.}
\def\physl #1 #2.{{\it Physics Letters} {\bf #1}, #2.}
\def\physrep #1 #2.{{\it Physics Reports} {\bf #1}, #2.}
\def\physrev #1 #2.{{\it Physical Review} {\bf #1}, #2.}
\def\physrevb #1 #2.{{\it Physical Review B} {\bf #1}, #2.}
\def\physrevd #1 #2.{{\it Physical Review D} {\bf #1}, #2.}
\def\physrevl #1 #2.{{\it Physical Review Letters} {\bf #1}, #2.}
\def\sovastr #1 #2.{{Soviet Astronomy} #1, #2.}
\def\sovastrl #1 #2.{{Soviet Astronomy (Letters)} #1, L#2.}
\def\commastr #1 #2.{{\it Comments Astrophys.} {\bf #1}, #2.}
\def\book #1 {{``{#1}''}}

\else

\def\aa #1 #2.{{Astr. \& Astrophys.} #1, #2.}
\def\araa #1 #2.{{Ann. Rev. Astr. \& %
Astrophys.} #1, #2.}
\def\aj #1 #2.{{A.J.} #1, #2.}
\def\apj #1 #2.{{Ap.J.} #1, #2.}

\def\apjl #1 #2.{{Ap.J. (Lett.)} #1, L#2.}

\def\apjsupp #1 #2.{{Ap.J. Supp.%
} #1, #2.}
\def\mnras #1 #2.{{M.N.R.\-a.S.} #1, #2.}

\def\qjras #1 #2.{{\it Q.J.R.\-a.S.} { #1}, #2.}
\def\nat #1 #2.{{ Nat.} { #1}, #2.}
\def\physl #1 #2.{{\it Phys. Lett.} {\bf #1}, #2.}
\def\physrep #1 #2.{{\it Phys. Rep.} {\bf #1}, #2.}
\def\physrev #1 #2.{{\it Phys. Rev.} {\bf #1}, #2.}
\def\physrevb #1 #2.{{\it Phys. Rev. B} {\bf #1}, #2.}
\def\physrevd #1 #2.{{\it Phys. Rev. D} {\bf #1}, #2.}
\def\physrevl #1 #2.{{\it Phys. Rev. Lett.} {\bf #1}, #2.}
\def\sovastr #1 #2.{{Sov. Astr.} #1, #2.}
\def\sovastrl #1 #2.{{Sov. Astr. (Lett.)} #1, L#2.}
\def\commastr #1 #2.{{\it Comm. Astr.} {\bf #1}, #2.}
\def\book #1 {{``{#1}''}}

\fi

%

\def\doublencite#1 #2{$^{\,\cref{#1},\cref{#2}}$}

\def\fromtoncite#1 #2{$^{\,\cref{#1}-\cref{#2}}$}

\def\nref#1{\penalty -100 \par \noindent \hangindent 1 truein 
\hbox to 0.5 truein {\hfil [\cref{#1}]}~}
%

\newif\ifapjnumbering
\def\apjnumbering{
	\nopagenumbers
	\headline={\ifnum\pageno > 1
			\hfil \folio \hfil
		   \else
			\hfil
			\fi}
                  }

\newif\ifsimboli
\newif\ifriferimenti

\newwrite\filerefs
\newwrite\fileeqs
\def\simboli{
    \immediate\write16{ !!! Genera il file \jobname.REFS }
    \simbolitrue\immediate\openout\filerefs=\jobname.refs
    \immediate\write16{ !!! Genera il file \jobname.EQS }
    \simbolitrue\immediate\openout\fileeqs=\jobname.eqs}
\newwrite\fileausiliario
\def\riferimentifuturi{
    \immediate\write16{ !!! Genera il file \jobname.AUX }
    \riferimentitrue\openin1 \jobname.aux
    \ifeof1\relax\else\closein1\relax\input\jobname.aux\fi
    \immediate\openout\fileausiliario=\jobname.aux}
%
%

\newcount\chapnum\global\chapnum=0
\newcount\sectnum\global\sectnum=0
\newcount\subsectnum\global\subsectnum=0
\newcount\subsubsectnum\global\subsubsectnum=0
\newcount\eqnum\global\eqnum=0
\newcount\citnum\global\citnum=0
\newcount\fignum\global\fignum=0
\newcount\tabnum\global\tabnum=0

\newcount\nnummore\global\nnummore=0

\def\itemore{\advance \nnummore by 1 \item{[\the\nnummore]}}

\newcount\nnummorerm\global\nnummorerm=0

\def\itemorerm{\advance \nnummorerm by 1 %
	\item{\expandafter{\romannumeral\the\nnummorerm})} }

\def\thechaproman{\uppercase\expandafter{\romannumeral \the \chapnum}}

\def\Chap#1{Chap.~\uppercase\expandafter{\romannumeral #1}}

\newif\ifndoppia
\def\numerazionedoppia{\ndoppiatrue\gdef\lasezionecorrente{
\the\sectnum }}

\def\seindefinito#1{\expandafter\ifx\csname#1\endcsname\relax}
\def\spoglia#1>{}

\def\cref#1{\seindefinito{@c@#1}\immediate\write16{ !!! \string\cref{#1}
    non definita !!!}
    \expandafter\xdef\csname@c@#1\endcsname{??}\fi\csname@c@#1\endcsname}

\def\eqref#1{\seindefinito{@eq@#1}\immediate\write16{ !!! \string\eqref{#1}
    non definita !!!}
    \expandafter\xdef\csname@eq@#1\endcsname{??}\fi\csname@eq@#1\endcsname}

\def\sectref#1{\seindefinito{@s@#1}\immediate\write16{ !!! \string\sectref{#1}
    non definita !!!}
    \expandafter\xdef\csname@s@#1\endcsname{??}\fi\csname@s@#1\endcsname}

\def\figref#1{\seindefinito{@f@#1}\immediate\write16{ !!! \string\figref{#1}
    non definita !!!}
    \expandafter\xdef\csname@f@#1\endcsname{??}\fi\csname@f@#1\endcsname}

\def\tabref#1{\seindefinito{@f@#1}\immediate\write16{ !!! \string\tabref{#1}
    non definita !!!}
    \expandafter\xdef\csname@f@#1\endcsname{??}\fi\csname@f@#1\endcsname}

%

\def\section#1\par{\immediate\write16{#1}\goodbreak\oneskip\halfskip
    \noindent{\bigbf #1}\nobreak\oneskip \nobreak\noindent}

\def\autosection#1#2\par { %
    \global\advance\sectnum by 1   %
 	\global\subsectnum=0
	\global\nnummore=0
	\global\nnummorerm=0
	\ifndoppia
		\global\eqnum=0
		\global\fignum=0
	    	\global\tabnum=0
		\fi
    \xdef\lasezionecorrente{\the\sectnum}
    \def\usaegetta{1}\seindefinito{@s@#1}\def\usaegetta{2}\fi

\expandafter\ifx\csname@s@#1\endcsname\lasezionecorrente\def\usaegetta{2}\fi
    \ifodd\usaegetta\immediate\write16
      { !!! possibili riferimenti errati a \string\sectref{#1} }\fi
    \expandafter\xdef\csname@s@#1\endcsname{\lasezionecorrente}
    \immediate\write16{\lasezionecorrente.#2}
    \ifsimboli
      \immediate\write\filerefs{ }\immediate\write\filerefs{ }
      \immediate\write\filerefs{  Sezione \lasezionecorrente : %
 sectref.   #1         page: \the\pageno}
      \immediate\write\filerefs{ }
      \immediate\write\fileeqs{ }\immediate\write\fileeqs{ }
      \immediate\write\fileeqs{  Sezione \lasezionecorrente : %
 sectref.   #1         page: \the\pageno}
      \immediate\write\fileeqs{ } \fi
    \ifriferimenti
      \immediate\write\fileausiliario{\string\expandafter\string\edef
      \string\csname@s@#1\string\endcsname{\lasezionecorrente}}\fi
\goodbreak \sectvskip %
\vtop{ %
\begingroup \hangindent 0.505truecm
	\noindent \hbox to  0.50truecm {{\bigbf\lasezionecorrente} \hfill}%
	{\bigbf #2} \par%
\endgroup
	}
	  \nobreak \vskip -\parskip \nobreak\indent 
	} %

\def\autosubsection#1#2\par { %
    \global\advance\subsectnum by 1
			\ifndoppia
				\global\eqnum=0
				\global\fignum=0
			    	\global\tabnum=0
			    	\global\subsubsectnum=0
				\global\nnummore=0
				\global\nnummorerm=0
				\fi
    \xdef\lasezionecorrente{\the\sectnum.\the\subsectnum}
    \def\usaegetta{1}\seindefinito{@s@#1}\def\usaegetta{2}\fi

\expandafter\ifx\csname@s@#1\endcsname\lasezionecorrente\def\usaegetta{2}\fi
    \ifodd\usaegetta\immediate\write16
      { !!! possibili riferimenti errati a \string\sectref{#1} }\fi
    \expandafter\xdef\csname@s@#1\endcsname{\lasezionecorrente}
    \immediate\write16{\lasezionecorrente.#2}
    \ifsimboli
      \immediate\write\filerefs{ } %
      \immediate\write\filerefs{  Sezione \lasezionecorrente : %
 sectref.   #1         page: \the\pageno}
      \immediate\write\filerefs{ }
      \immediate\write\fileeqs{ }  %
      \immediate\write\fileeqs{  Sezione \lasezionecorrente : %
 sectref.   #1         page: \the\pageno}
      \immediate\write\fileeqs{ } \fi
    \ifriferimenti
      \immediate\write\fileausiliario{\string\expandafter\string\edef
      \string\csname@s@#1\string\endcsname{\lasezionecorrente}}\fi
\goodbreak   
	\subsectvskip %
\vtop{ %
	\begingroup \hangindent 0.705truecm
	\noindent \hbox to  0.70truecm {{\it\lasezionecorrente} \hfill}%
	{\it #2} \par%
	\endgroup
	}
  \nobreak \vskip -\parskip \nobreak\indent
	} 

\def\autosubsubsection#1#2\par { %
    \global\advance\subsubsectnum by 1
				\global\nnummore=0
				\global\nnummorerm=0
    \xdef\lasezionecorrente{\the\sectnum.\the\subsectnum.\the\subsubsectnum}
    \def\usaegetta{1}\seindefinito{@s@#1}\def\usaegetta{2}\fi

\expandafter\ifx\csname@s@#1\endcsname\lasezionecorrente\def\usaegetta{2}\fi
    \ifodd\usaegetta\immediate\write16
      { !!! possibili riferimenti errati a \string\sectref{#1} }\fi
    \expandafter\xdef\csname@s@#1\endcsname{\lasezionecorrente}
    \immediate\write16{\lasezionecorrente.#2}
    \ifsimboli
      \immediate\write\filerefs{ }\immediate\write\filerefs{ }
      \immediate\write\filerefs{  Sezione \lasezionecorrente : %
 sectref.   #1         page: \the\pageno}
      \immediate\write\filerefs{ }
      \immediate\write\fileeqs{ }\immediate\write\fileeqs{ }
      \immediate\write\fileeqs{  Sezione \lasezionecorrente : %
 sectref.   #1         page: \the\pageno}
      \immediate\write\fileeqs{ } \fi
    \ifriferimenti
      \immediate\write\fileausiliario{\string\expandafter\string\edef
      \string\csname@s@#1\string\endcsname{\lasezionecorrente}}\fi
    \goodbreak  
	\subsectvskip  %
	\noindent \item{{\it\lasezionecorrente}}%
	{\it #2} \par %
	  \nobreak\halfskip \vskip -\parskip \nobreak\noindent
	} 

\def\semiautosection#1#2\par{
    \gdef\lasezionecorrente{#1}\ifndoppia\global\eqnum=0\fi
    \ifsimboli
      \immediate\write\filesimboli{ }\immediate\write\filesimboli{ }
      \immediate\write\filesimboli{  Sezione ** : sref.
          \expandafter\spoglia\meaning\lasezionecorrente}
      \immediate\write\filesimboli{ }\fi
    \section#2\par}

\def\eqlabel#1{\global\advance\eqnum by 1
    \ifndoppia\xdef\ilnumero{\lasezionecorrente.\the\eqnum}
       \else\xdef\ilnumero{\the\eqnum}\fi
    \def\usaegetta{1}\seindefinito{@eq@#1}\def\usaegetta{2}\fi
    \expandafter\ifx\csname@eq@#1\endcsname\ilnumero\def\usaegetta{2}\fi
    \ifodd\usaegetta\immediate\write16
       { !!! possibili riferimenti errati a \string\eqref{#1} }\fi
    \expandafter\xdef\csname@eq@#1\endcsname{\ilnumero}
    \ifndoppia
       \def\usaegetta{\expandafter\spoglia\meaning %
			\lasezionecorrente.\the\eqnum}
       \else\def\usaegetta{\the\eqnum}\fi
    \ifsimboli
       \immediate\write\fileeqs{     Equazione \ilnumero :%
  eqref.   #1           page: \the\pageno}\fi
    \ifriferimenti
       \immediate\write\fileausiliario{\string\expandafter\string\edef
       \string\csname@eq@#1\string\endcsname{\usaegetta}}\fi}

\def\autoeqno#1{\eqlabel{#1}
	\expandafter\eqno \hbox{%
	({\rm\ilnumero\kern 0.1ex})}
	}

\def\autoleqno#1{\eqlabel{#1}\leqno(\hbox{\rm \csname@eq@#1\endcsname})}

\def\eqmore{\global\advance\eqnum by 1
    \ifndoppia\xdef\ilnumero{\lasezionecorrente.\the\eqnum}
       \else\xdef\ilnumero{\the\eqnum}\fi
	\expandafter\eqno \hbox{%
	({\rm\ilnumero\kern 0.1ex})}
	}

\def\figlabel#1{\global\advance\fignum by 1
    \ifndoppia\xdef\ilnumero{\lasezionecorrente.\the\eqnum}
       \else\xdef\ilnumero{\the\fignum}\fi
    \def\usaegetta{1}\seindefinito{@f@#1}\def\usaegetta{2}\fi
    \expandafter\ifx\csname@f@#1\endcsname\ilnumero\def\usaegetta{2}\fi
    \ifodd\usaegetta\immediate\write16
       { !!! possibili riferimenti errati a \string\figref{#1} }\fi
    \expandafter\xdef\csname@f@#1\endcsname{\ilnumero}
    \ifndoppia
       \def\usaegetta{\expandafter\spoglia\meaning %
			\lasezionecorrente.\the\fignum}
       \else\def\usaegetta{\the\fignum}\fi
    \ifsimboli
       \immediate\write\fileeqs{                    Figure \figref{#1} : %
 figref.   #1               page: \the\pageno}\fi
    \ifriferimenti
       \immediate\write\fileausiliario{\string\expandafter\string\edef
       \string\csname@f@#1\string\endcsname{\usaegetta}}\fi}

%
%

\def\tablabel#1{\global\advance\tabnum by 1
    \ifndoppia\xdef\ilnumero{\lasezionecorrente.\the\tabnum}
       \else\xdef\ilnumero{\the\tabnum}\fi
    \def\usaegetta{1}\seindefinito{@f@#1}\def\usaegetta{2}\fi
    \expandafter\ifx\csname@f@#1\endcsname\ilnumero\def\usaegetta{2}\fi
    \ifodd\usaegetta\immediate\write16
       { !!! possibili riferimenti errati a \string\tabref{#1} }\fi
    \expandafter\xdef\csname@f@#1\endcsname{\ilnumero}
    \ifndoppia
       \def\usaegetta{\expandafter\spoglia\meaning  %
				\lasezionecorrente .\the\tabnum}
       \else\def\usaegetta{\the\tabnum}\fi
    \ifsimboli
       \immediate\write\fileeqs{                    Table \tabref{#1} : %
 tabref.   #1               page: \the\pageno}\fi
    \ifriferimenti
       \immediate\write\fileausiliario{\string\expandafter\string\edef
       \string\csname@f@#1\string\endcsname{\usaegetta}}\fi}

\def\clabel#1{\global\advance\citnum by 1%
  \xdef\lacitazione{\the\citnum}%
  \def\usaegetta{1}\seindefinito{@c@#1}\def\usaegetta{2}\fi%
  \expandafter\ifx\csname@c@#1\endcsname\lacitazione\def\usaegetta{2}\fi%
  \ifodd\usaegetta\immediate\write16%
       { !!! possibili riferimenti errati a \string\cref{#1} }\fi%
\expandafter\xdef\csname@c@#1\endcsname{\lacitazione}%
\ifsimboli%
  \immediate%
  \write\filerefs{   Citazione \lacitazione: #1  page: \the\pageno}\fi%
\ifriferimenti\immediate\write\fileausiliario%
    {\string\expandafter\string\edef\string\csname@c@#1\string\endcsname%
    {\lacitazione}}\fi%
 }%
%
%
\catcode`@ = 11
\newif\if@TestSubString
\def\IfSubString #1#2{%
	\edef\@MainString{#1}%
	\def\@TestSubS ##1#2##2\@Del{\edef\@TestTemp{##1}}%
		\expandafter\@TestSubS \@MainString#2\@Del
		\ifx\@MainString\@TestTemp
			\@TestSubStringfalse
		\else
			\@TestSubStringtrue
		\fi
		\if@TestSubString
		}
\def\Substitute@etal #1etal#2\@Del{\def\dummystring{#1\etal#2}}
\def\checketal#1{%
		\def\dummystring{#1}
		\IfSubString{#1}{etal}%
			\expandafter\Substitute@etal\dummystring\@Del%
		\fi%
\dummystring%
}
\catcode`@ = 12
\def\back3pt{\hbox{\kern -3pt}}
\def\cite #1/{\clabel{#1}\back3pt\checketal{#1}}
%
%
\catcode`@=11

\newdimen\@StrutSkip

\newdimen\@StrutSkipTemp

%
%
%
\def\SetStrut{%
   \@StrutSkip = \baselineskip
   \ifdim\baselineskip < 0pt
    \errhelp = {You probably called \string\offinterlineskip
		before \string\SetStrut}
     \errmessage{\string\SetStrut: negative \string\baselineskip
			(\the\baselineskip)}%
      \fi
        }

%
%
\def\MyStrut{%
	\vrule height 0.7\@StrutSkip
	        depth 0.3\@StrutSkip
 	        width 0pt
             }
%
%
%
\def \HigherStrut #1{%
     \@StrutSkipTemp = 0.7 \@StrutSkip
	\advance \@StrutSkipTemp by #1%
	\vrule height \@StrutSkipTemp depth 0.3\@StrutSkip width 0pt
      }

%
%
%
\def \DeeperStrut #1{%
     \@StrutSkipTemp = 0.3 \@StrutSkip
	\advance \@StrutSkipTemp by #1%
	\vrule height 0.7\@StrutSkip depth \@StrutSkipTemp width 0pt
      }
%
%
%
\def \LargerStrut #1{%
     \@StrutSkipTemp =  \@StrutSkip
	\advance \@StrutSkipTemp by #1%
	\vrule height 0.7\@StrutSkipTemp depth 0.3\@StrutSkipTemp width 0pt
      }

%
%
\SetStrut

%
%
%


\newcount\mscount
\def\multispan #1{%
      \omit
      \mscount = #1 %
      \loop \ifnum \mscount > 1
		\sp@n
	  \repeat
       }
\def\sp@n{\span\omit\advance\mscount by -1}

\catcode`@=12

%
%
%
%
\def\boxit#1{\leavevmode\hbox{\vrule\vtop{\vbox{\kern.33333pt\hrule
    \kern10pt\hbox{\kern10pt\vbox{#1}\kern10pt}}\kern10pt\hrule}\vrule}}

\magnification=1200
\def\parn{\par\noindent}

\overfullrule 0 pt
\baselineskip 6 true mm
\hsize 17 true cm
\nopagenumbers

\

\item \phantom {\bf {~~~~~~~~~~~~~STUDY OF THE CORE OF }}

\item \phantom {\bf {~~~~~~~~~~~~~THE SHAPLEY CONCENTRATION:}}

\item \phantom {\bf {~~~~~~~~~~~~~I. THE SAMPLE$^{\dagger}$}}

\

\item \phantom ~~~~Sandro BARDELLI$^{(1,2)}$,~Elena ZUCCA$^{(1,3)}$,
\item \phantom ~~~~Giampaolo VETTOLANI$^{(3)}$,~Giovanni ZAMORANI$^{(2,3)}$
\item \phantom ~~~~Roberto SCARAMELLA$^{(4)}$,~Chris A. COLLINS$^{(5)}$,
\item \phantom ~~~~and ~Harvey T. MacGILLIVRAY$^{(6)}$  .

\

\

\item \phantom ~~~~BAP 10--1993--030--DDA

\

\item \phantom {\it ~~~~The Monthly Notices of the Royal Astronomical Society
, in press.}

\

\item \phantom ~~~~Correspondence to S. Bardelli

\

\

\item \phantom ~~~~$^{(1)}$ Dipartimento di Astronomia, Universit\`a di
Bologna,
\item \phantom ~~~~via Zamboni 33, I--40126 Bologna, Italy

\

\item \phantom ~~~~$^{(2)}$ Osservatorio Astronomico di Bologna,
\item \phantom ~~~~via Zamboni 33, I-40126 Bologna, Italy

\

\item \phantom ~~~~$^{(3)}$ Istituto di Radioastronomia del CNR,
\item \phantom ~~~~via Irnerio 46, I-40126 Bologna, Italy

\

\item \phantom ~~~~$^{(4)}$ Osservatorio Astronomico di Roma,
\item \phantom ~~~~I--00040 Monteporzio Catone, Italy

\

\item \phantom ~~~~$^{(5)}$ Physics Departement, University of Durham,
\item \phantom ~~~~South Road, Durham, DH1 3LE, UK

\

\item \phantom ~~~~$^{(6)}$ Royal Observatory, Blackford Hill,
\item \phantom ~~~~Edinburgh, EH9 3HJ, UK

\vfill
\bigskip
\noindent
\item\phantom ${\dagger}$ Based on observations collected at the
European Southern Observatory, La Silla
\vfill
\eject
\pageno=1
\parn
{\centerline {\bf ABSTRACT}}
\par
We report the first results of a spectroscopic survey of galaxies in the
core of the Shapley Concentration, the richest nearby supercluster of clusters
of galaxies. We have measured 311 new galaxy redshifts in an area of $\sim 4.5$
square degrees centered around the Abell cluster {\it A3558}. Considering also
the data already available in the literature, the total number of galaxy
redshifts in this area amounts to more than 500.
\par
On the basis of these data we estimate the mean velocities and the velocity
dispersions of the Abell clusters {\it A3556},
{\it A3558} and the poor cluster {\it SC 1329 -314}. Finally, from an
analysis of the projected and three--dimensional distributions of galaxies in
this
region, we estimate the galaxy overdensity and find that the
core of the Shapley
Concentration has an interesting, very complex dynamical state: the main
clusters appear to be interacting with each other, forming a single elongated
structure containing many subcondensations.
\bigskip \bigskip
\parn
{\it Key words:} Galaxies: redshifts; Galaxies: clusters; Large--scale
structure.
\vfill\eject
\parn
{\bf 1. INTRODUCTION}
\par
The Shapley Concentration is the most remarkable feature (Scaramella et al.
1989; Vettolani et al. 1990)
which appears in studying the distribution of the Abell--ACO clusters of
galaxies
(Abell 1958; Abell, Corwin \& Olowin 1989).
Zucca et al. (1993) have analyzed, through a percolation algorithm,
superclusters of Abell--ACO clusters at various density excesses $f$ (defined
as the ratio between the local and the mean cluster density).
They found that at every density contrast the Shapley Concentration stands
out as the richest supercluster of the entire sky within a distance of
$300\ h^{-1}\ Mpc$ (hereafter $h = H_o / 100$); in particular, at $f\ge 2$
(which corresponds to a percolation radius of $\sim 25 \ h^{-1}\ Mpc$) it
has 25 members at a mean distance of $\sim 140 \ h^{-1}\ Mpc$
contained in a rectangular box of comoving sizes $\sim 32 \times
55 \times 100\ h^{-1}\ Mpc$ ($\alpha \times \delta \times\ distance$). The
central part of this concentration is already evident at high density excesses
($f\ge 100$, corresponding to a percolation radius of $\sim 6\ h^{-1}\ Mpc$)
and is formed by three condensations: {\it A3528--A3530--A3532}, {\it
A3571--A3572--A3575} and {\it A3556--A3558--A3560--A3562--A3564--A3566}.
The three clusters {\it A3556}, {\it A3558} and {\it A3562} (with the
poor cluster {\it SC\ 1329\ -314} in between {\it A3558} and {\it A3562}) form
an
aligned structure, elongated for
$\sim 3^o$ along the East--West direction and can be considered as the core of
the Shapley Concentration.
\par
The Shapley Concentration appears to be very rich and prominent also
studying the projected distribution of
optical galaxies (Raychaudhury 1989; Raychaudhury et al. 1991) and
the spatial distribution of both IRAS galaxies (Allen et al. 1990) and
X--ray clusters (Lahav et al. 1989). Indeed,
this region contains 6 of the 46 X--ray brightest clusters of the sky at
$|b^{II}|>20^o$ (Edge et al. 1990), {\it i.e.} $13 \%$ of the X--ray
brightest clusters reside in only $1.4 \%$ of the sky.
\par
Raychaudhury et al. (1991), assuming that the parameter $M^*$ of the optical
luminosity function of the clusters can be considered a standard candle,
concluded that
in this region there may be large deviations from the pure Hubble flow.
Moreover, on the basis of the available X--ray maps,  they noticed that the
fraction of multiple clusters in this
concentration is more than a factor 5 higher than in the ``field". Both these
facts suggest that this supercluster could be dynamically active, in the sense
that the processes like cluster evolution due to cluster--cluster or
cluster--group merging may be enhanced, as expected in an high density
environment.
\par
The Shapley Concentration is also likely to be an important player in
explaining the peculiar motion of the Local Group with respect to the Cosmic
Microwave Background frame. In fact, Scaramella et al. (1989, 1991)
pointed out that this supercluster may be responsible for a significant
fraction
($\simlt 30 \%$) of the Local Group peculiar motion, adding its dynamical pull
to that from a closer overdensity of galaxies at $\sim 40 \ h^{-1}\ Mpc$. The
latter
overdensity of galaxies, dubbed ``Great Attractor", was suggested to be a
major source of the Local Group acceleration (Lynden--Bell et al. 1988;
Lynden--Bell, Lahav \& Burstein 1989; Faber \& Burstein 1988; Dressler 1988).
The suggestions of Scaramella et al. (1989, 1991), on the contrary,
would imply a significantly larger coherence scale for the peculiar velocity
flow,
a fact which seems to be supported by recent findings (Willick 1990;
Mathewson, Ford \& Buchhorn 1992).
Also, Tully et al. (1992) suggested that these two ``attractors''
could be part of a single elongated planar structure, extending for
$\sim 450  \ h^{-1}\ Mpc$.
\par
The aim of this work is the study of the core of the Shapley Concentration,
with particular attention to the possible physical connection, made by
inter--cluster galaxies, between the main clusters.
In Sect.2 we present the sample and the new redshift data; in Sect.3 we discuss
the projected and three--dimensional distribution of the galaxies and in Sect.4
we
describe the methods we used to determine the mean velocity and the velocity
dispersion of the clusters and we apply them to each cluster. Finally, in
Sect.5 we summarize our results.
\bigskip
\parn
{\bf 2. THE SAMPLE}
\parn
{\bf 2.1 The galaxy catalogue}
\par
The photometric data catalogue is the COSMOS/UKST galaxy catalogue of the
southern sky (Yentis et al. 1992) obtained from the automated
scans of the UKST--J plates by the COSMOS machine.
The core of the Shapley
Concentration, as defined in Sect.1, is completely contained in the plate
\# 444, where 13600 galaxies are listed with $b_J \le 20$.
The catalogue lists, for each galaxy, accurate coordinates ($\alpha (2000)$
and $\delta(2000)$), the $b_J$ magnitude, the major diameter (in $arcmin$), the
area in pixels of the object, the ellipticity and the position angle.
{}From the catalogue
of plate 444 we have extracted a sub--sample of about $3^o.2 \times 1^o.4$ with
$13^h 22^m 06^s < \alpha < 13^h 37^m 15^s$ and $-32^o 22' 40''< \delta < -30^o
 59' 30''$,
which contains the complex {\it A3556}--{\it A3558}--{\it A3562}.
Fig. $1a$ shows isodensity contours for the galaxies with $b_J \leq 19.5$
in this area (2241 galaxies). Here we have chosen $19.5$ as the limiting
magnitude
for the contours since in our spectroscopic work (see Sect.2.2 and 2.3)
we have not observed galaxies fainter than this limit. The three Abell clusters
{\it A3558}, {\it A3556} and {\it A3562} are clearly visible in this figure,
as well as the poor cluster {\it SC 1329 -314}.
\par
In order to obtain three--dimensional information for the galaxy distribution
in this area, we had originally planned to cover it with a number of OPTOPUS
fields (Fig. $1b$), whose centers are listed in column
(2) and (3) of Table 1, together with the observation date, in column (4).
Three of these fields (\# 1, 4, 5) were planned to be observed twice, because
of their high density of galaxies. Unfortunately, observations of field \# 1
completely failed, for both technical and meteorological problems.
\parn
{\bf 2.2 Observations and data reduction}
\par
Spectroscopic observations were obtained at the ESO 3.6 m telescope in La
Silla,
equipped with the OPTOPUS multifiber spectrograph (Lund 1986), in the nights of
9--10--11 March 1991 and 25 April 1992.
The OPTOPUS multifiber spectrograph uses bundles of 50 optical
fibers, which can be set within the field of the Cassegrain focal plane
of the telescope; this field has a diameter of $32\ arcmin$, and each fiber
has a projected size on the sky of $2.5\ arcsec$.
We used the ESO grating $\#\ 15$
($300\ lines/mm$ and blaze angle of $4^o 18'$) allowing a dispersion of
$174$ \AA /mm in the wavelength range from $3700$ to $6024$ \AA.
The detector was the Tektronic $512 \times 512$  CCD with a pixel size of
$27\ \mu m$ corresponding to $4.5$ \AA, {\it i.e.} a velocity bin of
$\sim 270\ km / s $ at $5000$ \AA. We dedicated 5 fibers to sky measurements,
remaining with 45 fibers available for the objects. An average of about
3 spectra were lost in each exposure, due to broken or badly connected fibers.
 \par
The observing time for each field was one hour, split into two half--hour
exposures in order to minimize the effects due to the ``cosmic'' hits. The
observing sequence was: a 30 second exposure of a quartz--halogen white lamp,
a 60 second exposure of the Phillips Helium arc, then the first and the second
field exposures, and again the arc and the white lamp.
\par
The extraction of the one--dimensional spectra has been made using the
APEXTRACT
package as implemented in IRAF\footnote{$^{(1)}$}{IRAF is distributed by
KPNO, NOAO, operated by the AURA, Inc., for the National Science Foundation.}
. The loci of the spectra
were individuated and followed along the wavelength direction in the white
lamp frames and the solutions were used to find and extract the arcs and the
galaxy spectra.
 \par
The subtraction of the sky is the most critical step of the multifiber data
reduction, because it is impossible to separate sky and object in the same
spectrum (as instead done in the slit spectroscopy).
Moreover, the transmission of the light varies from fiber to fiber
and therefore it is not possible the direct subtraction of a sky
spectrum obtained with a fiber from a galaxy spectrum obtained with another
fiber.
In order to estimate the transmission of each fiber we adopted the following
procedure.
For each exposure, we fitted a Gaussian profile to the [OI] $\lambda$5577
sky line in each spectrum of the field and computed the continuum--subtracted
flux of this line. Under the assumption that the flux and the shape of the
spectrum of the night sky
do not appreciably vary inside the telescope field (this is true on scales
of $\sim$ degree, see also Wyse \& Gilmore 1992),
this flux should have the same value for all the spectra,
apart from a multiplicative factor, which is the transmission of each
fiber. In this way we could obtain an estimate of the relative transmission
of each fiber.
\par
In order to reveal possible contamination of the calibrating sky line
by ``cosmic''
hits, we computed also the mean and the dispersion of the ratios between the
peak of the continuum--subtracted counts of the [O I] $\lambda$5577 line and
the same quantity in the Na $\lambda$5891 sky line: values of this ratio
3 $\sigma$ above the mean have been taken to be indicative of the presence of
a ``cosmic'' hit in the [O I] $\lambda$5577 line.
However this contamination happened rarely and in all cases the presence of
the ``cosmic'' hit was clearly seen in a visual inspection of the spectra.
In these cases we eliminated the contamination by fitting the line below the
spike. After normalization for the relative fiber transmission,
the 5 sky spectra were averaged and subtracted from the object spectra.
The comparison of the two exposures of the same field revealed other
``cosmic'' spikes affecting the spectra, which were eliminated by interpolating
the continuum on both sides of the spikes.
\par
Radial velocities of galaxies whose spectra are dominated by absorption
features
have been determined using the program XCSAO of the IRAF task RVSAO, written by
Kurtz et al. (1992) following the cross--correlation
method of Tonry \& Davis (1979). After having fitted and subtracted
the continuum, we filtered
the spectra as suggested by Tonry \& Davis: low frequency cut--off
at the frequency 5, high frequency cut--off at 250 and full transmission
between 20 and 125. In determining the maximum value of the cross--correlation
function we fit the main peak above half maximum with a parabola.
As templates for the determination of the radial velocities we used the spectra
(kindly provided by L. Guzzo) of 6 galaxies, whose velocities have high
quality measurements in the literature.
Then, the ``best" velocity has been selected on the basis of the minimum
cross--correlation error rather than the maximum $R$ parameter of Tonry \&
Davis (1979). Indeed, being
$R$ the ratio between the height of the correlation peak and the r.m.s. of the
antisymmetric part of the correlation function, the choice of the maximum $R$
does not take into account the possible presence of very wide symmetrical
peaks. On the contrary, the errors, as calculated by Kurtz et al. (1992),
contain the FWHM of the peaks, called $w$, as
$$        error= {3 \over 8} { {{w} \over {(1+R)}}}  \eqno(1) $$
Finally, for spectra with strong emission line features we have used the EMSAO
program of the same IRAF task RVSAO.
\parn
{\bf 2.3 Redshift data }
\par
We have obtained a total of 383 spectra: 58 spectra ($\sim 15 \%$) were not
useful for redshift determination because of poor signal--to--noise ratio,
14 objects ($\sim 4 \%$ of the total number of spectra)
turned out to be stars, leaving us with 311 galaxy redshifts.
The median error on these velocities is $\sim 58\ km/s$.
\par
The galaxies of each observed field are given in Table 2, sorted by decreasing
magnitude.
Column (1) gives the identification number, column (2), (3) and (4) list the
right ascension, the declination and the $b_J$ magnitude, respectively.
Column (5) and (6) give the heliocentric velocity ($v=cz$) and its internal
({\it i.e.} cross--correlation) error; a blank in column (5) means that
the object was observed but it was impossible to derive velocity information
from its spectrum, while ``star" indicates those objects which
turned out to be stars (after the spectrum
analysis).
Finally, column (7) marks with the code $EMISS$ the redshifts determined
from emission lines.
\par
In Plates 1--8 we show the finding charts for all galaxies we observed.
\par
Our sample of spectroscopic data does not contain any galaxy with already
published velocities. There are, however, 45 galaxies in common with
those observed by
Stein et al. (private communication). Being these two sets of data
taken with the same instrument but reduced and cross--correlated with
different packages (IRAF vs MIDAS), a
direct comparison of the velocities gives a rough estimate of the
external errors in the data. We have compared our velocities ($v_{us}$) with
those of Stein et al. ($v_{Stein}$), taking into account the
errors ($err_{us}$ and $err_{Stein}$).
 Computing the mean and the {\it r.m.s.} of the variable
$$    {{v_{Stein} - v_{us}} \over { \sqrt{ err_{Stein}^2 + err_{us}^2 }}}
     \eqno(2) $$
we obtain $mean= 0.23 \pm 0.28$ and $r.m.s.=1.87$, to be compared with
the expected values of 0 and 1, respectively. The resulting mean shows
that the two sets of measurements are consistent with having the same velocity
zero point, while the value of the {\it r.m.s.} has a probability lower than
0.001 (through a $\chi^2$ test) to be compatible with the value 1.
Thus, assuming that for both sets of data the quoted errors are good estimates
of the cross--correlation errors, the value of
the {\it r.m.s.} suggests that the true statistical errors are on average
$1.87$ times greater than the cross--correlation ones. This result is
consistent
with a similar analysis performed by Malumuth et al. (1992): using
multiple observations of 42 galaxies, reduced in the same way, they concluded
that this factor is of the order of $1.6$.
\par
Adding to our data additional redshifts available in the literature, the total
number of redshifts in the area shown in Fig. 1 is 511, 200 of which are
taken from the literature and 311 are new measurements.
\bigskip
\parn
{\bf 3. VELOCITY DISTRIBUTION AND OVERDENSITY}
\par
Fig. 2 shows velocity versus apparent magnitude for all the galaxies
with measured redshifts in our sample. In addition to the extremely prominent
structure at $v\sim 15000\ km/s$, other clumps of galaxies at $v\sim 25000,
\ 40000$ and $55000\ km/s$ are visible, separated by ``voids" of typical
size of $\sim 5000\ km/s$. In the foreground of the Shapley Concentration
there are a few galaxies with $v\sim 4000\ km/s$, consistent with the
velocity of the Great Attractor.
\par
The significance of these voids and peaks is not easy to be determined, because
no well defined magnitude completeness limit can be determined for our
spectroscopic data. Table 3 shows, as a function of magnitude, the total
number of objects in the photometric catalogue, the number of galaxies
with measured redshift (including also literature data), the number of stars
and the percentage of spectroscopic data in the area covered by our OPTOPUS
fields. As it is clear from the last column of the table, the completeness of
our spectroscopic data is smoothly decreasing with magnitude, from $\sim
75 \%$ for the brightest galaxies down to $\sim 15 \%$ for the faintest
galaxies in the sample.
\par
{}From these data we have estimated the expected distribution
of uniformly distributed galaxies by integrating a luminosity function with
parameters $\alpha=-1.10$ and $M^* = -19.67 + 5 \log h$ (Vettolani et al. 1993)
and weighting the expected number of galaxies in each magnitude bin by
the corresponding spectroscopic completeness.
The normalization has been adjusted in such a way as
to reproduce the observed number of galaxies outside the velocity range of the
Shapley Concentration. To this purpose we have adopted the velocity intervals
$v<11000 \ km/s$ and $v>18000\ km/s$. We have also verified that the
integration
of the luminosity function with the here adopted normalization reproduces
reasonably well the counts of galaxies with $b_J \simlt 19.5$
(Heydon--Dumbleton,
Collins \& MacGillivray 1989). Fig. $3a$ shows the histogram of the
observed velocities (in bins of $1000\ km/s$) and the resulting expected
distribution. Note that, in order to visualize the distribution outside the
Shapley Concentration, the Y--axis has been cut at $N=20$, while the bin
$14000 <v< 15000\ km/s$ contains 166 galaxies.
\par
This histogram suggests the possible presence of a void of about $4000\ km/s$
just behind the main peak. No galaxy is seen in the velocity range
$18000 - 22000\ km/s$, where $\sim 10$ galaxies would be expected on the basis
of a uniform distribution. Other possible voids, with approximately similar
statistical significance, are seen in the velocity ranges $28000 - 34000\
km/s$ and $44000 - 51000\ km/s$.
\par
Fig. $3b$ shows a close--up of the distribution of the observed velocities in
the range $10000 - 20000\ km/s$; note that the velocity bin here is
$500\ km/s$. The average velocity
is $ 14248\ km/s$, with a Gaussian velocity dispersion of $1083\ km/s$: these
are the parameters of the Gaussian curve superimposed on the velocity
distribution.
\par
To estimate the overdensity of galaxies in this area is not straightforward,
because the observed velocity distribution is the convolution of the distance
distribution with the velocity dispersion within the clusters. As such, we
can not unambiguously compute the depth distribution of the galaxies. The
total number of galaxies in the velocity range $11000 -18000\ km/s$ is 430, to
be compared with an expected number of $\sim 15$, on the basis of a uniform
distribution. This would correspond to an overdensity $ (N - N_{exp})/N_{exp}
\simeq 28$, over a depth of $70\ h^{-1}\ Mpc$. Therefore, this estimate is a
lower
limit to the real spatial overdensity of the bulk of the galaxies in this
region.
\par
A second, more stringent, lower limit to the overdensity for the clusters
in the core of the Shapley Concentration can be obtained by comparing
the number of galaxies within $\pm 1\sigma$ from the average velocity
($N=304$) with the expected number of galaxies within the same velocity
range ($N_{exp}\sim 5$). From this we derive an estimate of an overdensity
of $\sim 65$ over a depth of $\sim 20\ h^{-1}\ Mpc$.
\par
It is important to note here that these overdensities have been estimated
with respect to an assumed uniform background and do not take into
account the possibility that the area analyzed here is the peak of a much
more extended overdensity, not simply associated to Abell clusters. That
this might be the case is shown in Fig. 4, which represents the ratio
between the differential galaxy counts as a function of magnitude for the whole
plate 444 with respect to those of a reference area in a region of $\sim 140$
square degrees, near the South Galactic Pole (filled circles). Open
circles show the same ratio after excluding all the galaxies
within 1 Abell radius of the 5 Abell clusters of the Shapley Concentration
lying in this plate (Zucca et al. 1993). At each magnitude the surface
density of galaxies in plate 444 is significantly higher than in the reference
area, even after excluding the cluster galaxies. In particular, the observed
surface density of inter--cluster galaxies is about twice that the reference
area at $b_J \sim 16.2$, corresponding to $M^*$ galaxies at a distance of
$\sim 140\ h^{-1}\ Mpc$. A more detailed study of the velocity distribution
and the spatial overdensity of all galaxies with $b_J \le 16.5$ in the
entire plate 444 is in progress (Schuecker et al., in preparation).
\par
In Fig. 5 we show the wedge diagram of the galaxies of our sample in the
velocity range $10000 - 24000\ km/s$. The three pairs of straight lines
(solid, dashed and dotted) show the projection in right ascension of 1 Abell
radius for the three clusters {\it A3562}, {\it A3558} and {\it A3556},
respectively. Apart from the ``hole" due to the lack of data in the area of
the failed field \# 1, centered on the cluster {\it A3562}, the galaxy
distribution seems to form a single connected structure. However, the mean
velocity of
{\it A3562} as reported by Melnick \& Moles (1987), $15060 \ km/s$, suggests a
continuity also across the failed field. The main features in this figure are:
\parn
{\bf a)} the ``Fingers of God" visible at the center of the two clusters
{\it A3558} and {\it A3556};
\parn
{\bf b)} two groups of galaxies, clearly separated
in velocity, at about $v\sim 14400\ km/s$ and $v\sim 12200 \ km/s$ eastward of
{\it A3562};
\parn
{\bf c)} two possibly separated groups at $v\sim 13500\ km/s$ and $v\sim 15000\
km/s$ at $\alpha \sim 13^h 30^m$, corresponding to about 1 Abell radius
from both {\it A3558} and {\it A3562} (see also the discussion below). Note
that this right ascension corresponds to the approximate position of the
density enhancement identified with the poor cluster {\it SC 1329 -314};
\parn
{\bf d)} a zone devoid of galaxies just behind the main structure,
extending radially from $17250$ to $22000\ km/s$ (see the discussion above
in this section).
\par
In order to better understand how these features are related to the
projected distribution of galaxies, we have superimposed the galaxies
with redshift information to isodensity contour maps of the region.
Fig. 6 shows the most interesting part of these plots, centered on {\it A3558}:
panel $(a)$ refers to all galaxies with $b_J \le 19.5$, while panels $(b)$,
$(c)$ and $(d)$ show the isodensity contours for galaxies with $b_J \le 18$,
$18 < b_J \le 19$ and $19 < b_J \le 19.5$, respectively.
The isodensity levels have been chosen as integer multiple of the mean number
of galaxies per pixel in the entire plate (the values for each panel are
given in the figure captions). The size of the superimposed dots is inversely
proportional to the velocity of the galaxies. The main features of Fig. $6a$
are the density peak of {\it A3558} (in the center) extending on the East
toward
the poor cluster {\it SC 1329 -314} and the overdensity corresponding to
{\it A3556} on the West: note that {\it A3558} and {\it A3556} appear to be
unconnected only due to our choice of the lower isodensity value.
{\it A3558} appears to be elongated with the major axis at a position angle
of $\sim 135^o$ (measured from North toward East): such a departure from
circular symmetry, with a similar position angle, is seen also in ROSAT
X--ray maps of this cluster (Bardelli et al., in preparation).
\par
Fig. $6a$ is also characterized by a number of smaller substructures: in
particular it is clear the presence of two subcondensations (arrowed in the
figure) at different
velocities at the approximate position of the poor cluster {\it SC 1329 -314}
(see Sect.4.3). Galaxies in the subcondensation $A$ have,
on average, smaller velocity (higher fraction of big dots) than galaxies in
the subcondensation $B$. The galaxies in these two clumps appear to have also
a different magnitude distribution. While subcondensation $A$ is clearly
visible on the contour plot at brighter magnitude, the opposite is true for
subcondensation $B$. This effect would be expected if $B$ were at a
significantly larger distance than $A$, contrary to what is suggested by
the velocity data ($\Delta v \sim 750\ km/s$, see Sect.4.3). The conclusion
would be that either the galaxies in the two subcondensations have quite
different luminosity function or there are strong peculiar motions which
can alter significantly the correspondence between velocity and distance.
An other interesting feature which is clearly seen in the maps of Fig. 6
is the decrease with magnitude of the contrast between the surface density
of the center of {\it A3558} and the average density over the entire plate.
While this contrast is $\sim 15$ for galaxies with $b_J \le 18$, it
decreases to $\sim 3$ for galaxies with $19 < b_J < 19.5$. In the latter
magnitude range the center of {\it A3558} is not more conspicuous than
subcondensation $B$.
\par
The presence of these substructures makes the dynamical situation of this
region quite complex. For this reason, the estimate of the masses of these
clusters is not straightforward and needs a careful analysis of the
contribution of the various subcondensations: estimates for the masses, the
luminosity functions and
the mass--to--light ratios for these clusters will appear elsewhere (Bardelli
et al., in preparation).
\par
\bigskip
\parn
{\bf 4. MEAN VELOCITY AND VELOCITY DISPERSION}
\par
{\bf OF THE CLUSTERS}
\par
In order to estimate the mean velocity and the velocity dispersion for the
three clusters {\it A3558}, {\it A3556} and {\it SC 1329 -314} we have used
the biweight location ($C_{BI}$) and scale ($S_{BI}$) estimators discussed by
Beers, Flynn \& Gebhardt (1990) and already adopted for the study of {\it
A3558}
by Teague, Carter \& Gray (1990). These estimators are defined as:
$$ C_{BI}= {\rm m} + {{\sum_{|u_i|<1} (v_i - {\rm m}) (1 - u_i^2)^2}
\over { \sum_{|u_i|<1} (1 - u_i^2)^2}} \eqno(3) $$
$$ S_{BI} = \sqrt {N}\ { \left[{\sum_{|u_i|<1} {(v_i - {\rm m})^2
(1 - u_i^2)^4}}\right] ^{1/2}
\over { |\sum_{|u_i|<1} (1 - u_i^2) (1-5 u_i^2)|} }  \eqno(4) $$
$$u_i= {{v_i - {\rm m}} \over {c \cdot median(|v_i - {\rm m}|)}}  \eqno(5) $$
where ${\rm m}$ is the median of the data, $ median(|v_i - {\rm m}|)$ is the
median value of the variable $(|v_i - {\rm m}|)$ and the ``tuning constant"
$c$, following Beers et al., has been set equal to 6.0 for the location
estimator and equal to 9.0 for the scale estimator.
The physical meaning of these estimators is analogous to that of the
mean ($<v>$) and the standard deviation
($\sigma_v$), to which they asymptotically approach when the sample is taken
from a Gaussian distribution, but they are proved to be more robust and
resistant than the classical estimators (Beers et al. 1990).
\par
These estimators are characterized by a cut--off on the data tails
corresponding, for a Gaussian distribution, to $\sim 4 \sigma$
for $C_{BI}$ and to $\sim 6 \sigma$ for $S_{BI}$ and a decreasing weight
to the most extreme data. In order to calculate these quantities, we have used
the program ROSTAT, kindly given us by T. Beers; the errors are the $68 \%$
uncertainties computed after one hundred bootstrap resampling of the data.
\par
In order to find the velocity range in which the cluster members lie, we have
assumed that the velocity distribution of cluster galaxies is Gaussian, as
expected when the system has undergone a violent relaxation. In order to
individuate this velocity range, we have applied the following iterative
procedure. First, we have computed $C_{BI}$ and $S_{BI}$ using all galaxies,
without limits in velocity, and we have checked the Gaussianity.
In order to check the hypothesis of Gaussianity, we performed 4
standard tests: the $a$ test (defined as the ratio between the mean deviation
and the standard deviation), the $b_1$ (skewness) test, the $b_2$ (kurtosis)
test and the $I$ test (the ratio between the standard deviation estimator and
the biweight scale estimator).
The definitions and the percentage points for $a$,
$b_1$ and $b_2$ are reported in Pearson \& Hartley (1962), while the $I$ test
and the formulae to calculate the critical points are reported by
Teague et al. (1990). If one or more tests rejected the null hypothesis
of Gaussianity, following a suggestion in Beers et al. (1990) we have
repeated the analysis using the velocity range $ C_{BI} \pm 6\cdot
median(|v_i - {\rm m}|)$ and we have checked again the Gaussianity of the new
velocity distribution. We have iterated this procedure until the velocity
distribution of galaxies was consistent with being Gaussian at more than $5\%$
probability level for all the four tests. In all cases convergence has been
reached in a few iterations.
\par
The velocity histograms for the three clusters are shown in the three panels
of Fig. 7, where the adopted radii are $36\ arcmin$ ($=1$ Abell radius) for
{\it A3558}, $18\ arcmin$
for {\it A3556} and $6\ arcmin$ centered at the positions of the two
subcondensations $A$ and $B$ (see previous section) for {\it SC 1329 -314}.
In the last panel the dashed histogram refers to subcondensation $A$, while
the solid one refers to subcondensation $B$.
\parn
{\bf 4.1 Abell 3558 }
\par
{\it A3558}, also known as Shapley 8 or {\it SC\ 1325\ -311}, is the richest
cluster contained in the ACO catalogue, the only one with richness class
4. It is dominated by a central giant galaxy and it is a strong X--ray emitting
cluster with an estimated luminosity of $8.7\times 10^{44}\ ergs/s $,
in the energy range $[2 - 10]\ keV$ (Day et al. 1991). Moreover, this
object is also interesting because Gebhardt \& Beers (1991) found that it is
one of the 4 clusters which are claimed to present a statistically significant
discrepancy (at $90\%$ level) between the velocity of the centroid of the
redshifts and the velocity of the dominant galaxy.
\par
Within a circle of 1 Abell radius ($\sim 1.5\ h^{-1} Mpc$) around this cluster
we have 267 galaxies with measured redshifts in the velocity range
$10260 - 18516\ km/s$ (chosen following the iterative procedure described in
the previous section), out of which $\sim 100$ are taken from Metcalfe, Godwin
\& Spenser (1987) and Teague et al. (1990); the velocity histogram of these
galaxies is shown in Fig. $7a$.
Because of the large number of redshift data now available we have divided the
total sample in 6 sub--samples at different distances from the center and
have analyzed them separately. Table 4 lists the main properties of these
sub--samples. Column (1) and (2) give the distance from the center in $arcsec$
and $h^{-1}\ Mpc$ respectively, column (3) and (4) list the number of galaxies
in our bi--dimensional sample and the number of measured redshifts in the
velocity range $10260 - 18516 \ km/s$.
\par
In order to better understand the contribution to $C_{BI}$ and $S_{BI}$ from
the
galaxies at different distances from the cluster center, we have analyzed not
only the ``integral" sub--samples described above, but also ``differential"
sub--samples, selected as annuli around the cluster center, with
distances in the ranges $0-360$, $360-718$, $718-1077$, $1077-1435$,
$1435-1794$ and $1794-2155\ arcsec$, respectively. The four panels of Fig. 8
show the mean velocity and the velocity dispersion for the integral
(panels $a$ and $c$) and the differential (panels $b$ and $d$) sub--samples,
whose values are reported in Table 5.
The $C_{BI}$ and $S_{BI}$ derived for the integral sub--samples are
approximately constant up to half an Abell radius, and in good agreement with
previous estimates: indeed, we find (for
the third sub--sample) $C_{BI}=14242\pm 80\ km/s$ and $S_{BI}=986\pm 60\ km/s$,
while Metcalfe et al. (1987) report $<v>=14237\ km/s$ and $\sigma=
991\pm 157\ km/s$ and Teague et al. (1990) give $<v>=14233\ km/s$ and
$\sigma=1002\ km/s$.
\par
However, there is a tendency for $C_{BI}$ to increase at larger
distances from the center: this feature is very clear for the differential
sub--samples (Fig. $8b$), where there is a significant discontinuity between
the third and the fourth sub--samples. This behaviour is probably a consequence
of the fact that, as the distance from the center increases, the contamination
by galaxies belonging to subcondensation $B$ of the poor cluster {\it SC 1329
-314} becomes higher.
The distribution of $S_{BI}$ in the differential sub--samples (Fig. $8d$) does
not show any statistically significant trend.
\par
As already mentioned above, Gebhardt \& Beers (1991) noticed that {\it A3558}
(and other three clusters) presents a significant offset between the mean
velocity of the cluster and the velocity of the dominant galaxy ($v_D =
14037\pm 21$; Teague et al. 1990). Our data, taken at face value, seems
to confirm such a discrepancy. In fact, limiting ourselves at the first three
integral sub--samples, we find that the difference between $C_{BI}$ and $v_D$
is significant at $\sim$ 1.6, 1.8 and 2.5 $\sigma_D$ level respectively,
where
$$ \sigma_D = {{| C_{BI} - v_D |} \over
	      {\sqrt{ err^2 _{C_{BI}} + err^2 _{v_D} } }} \eqno(6) $$
The trend of increasing significance with increasing distance from the center
is not due to changes of $C_{BI}$, but rather to the decrease of $err_{C_{BI}}$
when more galaxies are included in the sample. However, in the light of the
possibility that the contamination on $C_{BI}$ from different subcondensations
might not be negligible, even in the inner region of the cluster, we have
repeated the same analysis by dividing each sub--sample in two parts,
eastward and westward of the center. The result is that the mean velocities
of the three West sub--samples are all inconsistent with $v_D$ at more than
$2\sigma_D$ level, while the discrepancy is reduced to 0.28, 0.03, 0.93
$\sigma_D$ for the first three East sub--samples. This result suggests
that the observed discrepancy between $C_{BI}$ and $v_D$ is at least partly,
if not completely, due to asymmetries in the velocity distribution of the
galaxies in the cluster.
\parn
{\bf 4.2 Abell 3556 }
\par
{\it A3556} is an ACO cluster of richness class 0 and Bautz--Morgan type I.
Its angular distance from {\it A3558} is only $\sim 50\ arcmin$ and there is
an overlap between the Abell radii of these two clusters (see Fig. $1a$).
For this reason, we have analyzed for this cluster the velocity distribution
of the galaxies within half an Abell radius and in the velocity range
$11293 - 17509\ km/s$, chosen following the Gaussianity criterium (see Fig.
$7b$). From this sample of 48 galaxies we obtain $C_{BI} = 14407 \pm 89\ km/s$
and $S_{BI} = 554 \pm 47\ km/s$. The average velocity $C_{BI}$ of {\it A3556}
is therefore consistent (at $1.4\sigma$ level) whit that of {\it A3558}.
\parn
{\bf 4.3 The poor cluster SC 1329 -314 }
\par
The poor clusters {\it SC 1329 -314} is in between {\it A3558} and
{\it A3562}. On the basis of 11 redshifts, Melnick \& Moles (1987) derived
for this cluster a mean velocity of $13300\ km/s$ and a velocity dispersion
of $1050\ km/s$, but they do not give the positions of their measured galaxies.
As discussed above, our data suggest the presence of at least two possibly
separated subcondensations in this area. We have determined $C_{BI}$ and
$S_{BI}$ separately in these two substructures, using all galaxies within
$6\ arcmin$ around tentative center positions estimated from the surface
density contour plot.
We found $C_{BI} (A) = 14074 \pm 249\ km/s$ and $S_{BI} (A) = 1044 \pm 96
\ km/s$ for the 28 galaxies in $A$ and $C_{BI} (B) = 14828 \pm 225\ km/s$ and
$S_{BI} (B) = 676 \pm 189 \ km/s$ for the 14 galaxies in $B$ (see Fig. $7c$).
Subcondensation $A$ consists of brighter galaxies (see Fig. $6b$) and is
closer than $B$ to the nominal center of the cluster: therefore it is likely
that Melnick \& Moles (1987) measured galaxies mainly in this group, thus
obtaining a lower value for the mean velocity.
\par
Finally, we note the presence of two peaks of X--ray emission probably
associated with these subcondensations. In our ROSAT PSPC image pointed on
{\it A3558} it is present, in addition to the cluster, a secondary peak of
X--ray emission at a distance of a few arcminutes from subcondensation $B$,
and another peak in an image taken by the Einstein Observatory
(as reported by Raychaudhury et al. 1991) is very near to subcondensation $A$.
This correspondence seems to confirm the physical reality of these two
associations of galaxies.
\bigskip
\parn
{\bf 5. SUMMARY }
\par
In this paper we presented and discussed a sample of galaxies in the core of
the Shapley Concentration, formed by $\sim 500$ galaxy redshifts (311 of which
are new measurements) in a region of the sky included in a rectangle of
$\sim 3^o.2 \times 1^o.4$.
The new redshifts were obtained with multifiber spectroscopy and the
median error on the velocities is $\sim 58\ km/s$.
\par
{}From an analysis of the spatial distribution of this sample, combined with
the bi--dimensional isodensity contours of the COSMOS/UKST catalogue, we
conclude that the core of the Shapley Concentration is in a complex dynamical
state. The three main clusters of this region ({\it A3558}, {\it A3556} and
{\it A3562}) are each other interacting and form a single elongated structure,
orthogonal to the line of sight,
with many subcondensations. In particular, we have studied two
subcondensations around the position of
the poor cluster {\it SC 1329 -314}, which is in between {\it A3558}
and {\it A3562}. These substructures are probably
real groups of galaxies, subjected to the gravitational fields of the main
clusters.
\par
The galaxy overdensity in this area, estimated with respect to an average
uniform background, is $\simgt 65$. Being a lower limit, this number is
consistent
with the overdensity of the order of $110 \pm 40$ found by Postman {\it
et al.} (1988) for the galaxies in the clusters of the Corona Borealis
supercluster. From an analysis of the galaxy counts in the whole plate 444
we conclude that such an overdensity is present, although at a much lower
level, also for inter--cluster galaxies over a much larger area.
\par
Furthermore, we have derived the mean velocity and the velocity dispersion for
{\it A3558}, {\it A3556} and {\it SC 1329 -314}, using the biweight estimator
of the location and scale. In particular, for {\it A3558} we have calculated
these quantities as a function of the distance from the cluster center,
concluding that also in deriving these parameters the contamination from the
subcondensations is not negligible, expecially at distances greater than
0.5 Abell radii.
\par
{}From this analysis it is clear that the core of the Shapley Concentration is
dynamically very active: therefore it is very interesting to derive the
masses and the mass--to--light ratios of these clusters, as well as the
luminosity function of the galaxies, in order to estimate the time scale of
their merging. We will present elsewhere (Bardelli et al., in
preparation) the results of this dynamical analysis.
\vskip 3 true cm
\parn
{\bf Acknowledgements:}
\parn
We warmly thank Paul Stein for having allowed us the use of his redshifts in
advance of the publication, T.C. Beers for having kindly given us his program
ROSTAT, Luigi Guzzo for the template spectra and R.Primavera for the
photographic artwork. This work has been partially supported by the
Italian Space Agency (ASI) under the contract ASI 91--RS--86.
\parn
\vfill\eject
{\centerline {\bf FIGURE CAPTIONS }}
 \bigskip
\bigskip
\parn
{\bf Figure 1:}
\parn
{\it a)} Isodensity contours of the core of the Shapley Concentration
in an area of $\sim 3^o.2 \times 1^o.4$.
The figure refers to galaxies with $b_J \le 19.5$ and binned in
$ 2\ arcmin \times 2\ arcmin$ bins; the data have been smoothed with a Gaussian
with a FWHM of $6\ arcmin$. For the three Abell clusters circles
of one Abell radius have been drawn (dashed curves); the poor cluster
{\it SC 1329-314} is the peak between the clusters
{\it A3558} and {\it A3562}.
\parn
{\it b)} Same as Fig. $1a$, with superimposed the nine OPTOPUS fields
observed in March 1991 and April 1992.
\parn
{\bf Figure 2:}
\parn
Velocity versus apparent magnitude for all the galaxies with measured
redshifts in our sample.
\parn
{\bf Figure 3:}
\parn
{\it a)} Histogram of the observed velocities in bins of $1000\ km/s$ for
galaxies with $b_J \le 19.5$. The superimposed dashed curve corresponds to
the distribution expected for uniformly distributed galaxies. Note that
the Y--axis has been cut at $N=20$ in order to visualize the distribution
outside the Shapley Concentration.
\parn
{\it b)} Close--up of the distribution of the observed velocities in the
range $10000 - 20000\ km/s$; the velocity bin is $500\ km/s$ and the
limiting magnitude is $b_J = 19.5$. The solid curve is a Gaussian with
$<v>=14248\ km/s$ and $\sigma=1083\ km/s$.
\parn
{\bf Figure 4:}
\parn
Differential counts as a function of magnitude normalized to
those of a reference area near the South Galactic Pole. Solid circles: whole
plate 444; open circles: plate 444 without galaxies within 1 Abell radius
from the center of the clusters of the Shapley Concentration. The counts of
plate 444 have been corrected for galactic absorption. For clearness, error
bars (at $1\sigma$) have been drawn only for the solid circles; the errors for
the open
circles are comparable. Error bars for $b_J > 17.5$ are smaller than the
symbol size.
\parn
{\bf Figure 5:}
\parn
Wedge diagram of the sample of galaxies in the velocity range $10000 -
24000\ km/s$. The coordinate range is $13^h 22^m 06^s < \alpha (2000)< 13^h
37^m
15^s$ and $-32^o 22' 40"< \delta(2000) < -30^o 59' 30"$.
The straight lines drawn in the picture show the projection in right ascension
of a circle of 1 Abell radius for each cluster: solid lines refer to
{\it A3562}, dashed and dotted lines to {\it A3558} and {\it A3556},
respectively.
\parn
{\bf Figure 6:}
\parn
Isodensity contours of the galaxies in an area of $\sim 2^o \times 2^o$ around
{\it A3558}. The contours refer to galaxies binned in $ 1\ arcmin \times
1\ arcmin$ bins; the data have been smoothed with a Gaussian with a FWHM of $6
\ arcmin$. The isodensity levels for each magnitude range have been chosen as
integer multiple of the mean number of galaxies per pixel ($<n>$) in the
entire plate. The dots represent galaxies with measured redshifts in the
velocity range $11250 - 18000\ km/s$ and their size is inversely proportional
to the radial velocity (the larger the dot the smaller the velocity).
The four different sizes for the dots correspond to the velocity ranges
$11250 - 13250$, $13250 - 14250$, $14250 - 15500$ and $15500 - 18000\
km/s$.
\parn
{\it a)} Galaxies with $b_J \le 19.5$; the isodensity levels are $2<n>$,
$3<n>$, $4<n>$, $5<n>$, $6<n>$, and $7<n>$, where $<n>=0.09$. The two
dashed straight lines indicate subcondensations $A$ and $B$, discussed in
the text.
\parn
{\it b)} Galaxies with $b_J \le 18.0$; the isodensity levels are $2<n>$,
$4<n>$, $6<n>$, $8<n>$, $10<n>$, $12<n>$ and $14<n>$, where $<n>=0.02$.
\parn
{\it c)} Galaxies with $18.0< b_J \le 19.0$; the isodensity levels are $2<n>$,
$4<n>$, $6<n>$, and $8<n>$, where $<n>=0.03$.
\parn
{\it d)} Galaxies with $19.0 < b_J \le 19.5$; the isodensity levels are $2<n>$,
$3<n>$ and $4<n>$, where $<n>=0.04$.
\parn
{\bf Figure 7:}
\parn
Velocity histogram of the galaxies within {\it a)} 1 Abell radius around
{\it A3558}, {\it b)} 0.5 Abell radii around {\it A3556} and {\it c)}
$6\ arcmin$ around the positions of subcondensations $A$ (dashed histogram) and
$B$ (solid histogram) in the poor cluster {\it SC 1329 -314}. The velocity
ranges for galaxies in panel {\it a)} and {\it b)} have been chosen following
the Gaussianity criterium (see text for more details).
\parn
{\bf Figure 8:}
\parn
Mean velocity ($C_{BI}$) and velocity dispersion ($S_{BI}$) of {\it A3558}
as function of the angular distance from the cluster center. Panels {\it a)}
and {\it c)} refer to the ``integral" samples, while panels {\it b)} and
{\it d)} refer to ``differential" samples (see Sect.4.1).
\bigskip
{\centerline {\bf PLATE CAPTIONS }}
 \bigskip
\parn
{\bf Plate 1:}
\parn
Finding chart for the galaxies with new redshift data in the OPTOPUS field
$\# 2$. Reproduced from the ESO Sky Survey
R plates; North is at the top, East at the right. Numbers correspond to the
identification number of the galaxies as in Table 2.
\parn
{\bf Plate 2:}
\parn
Finding chart for the galaxies with new redshift data in the OPTOPUS field
$\# 3$. Reproduced from the ESO Sky Survey
R plates; North is at the top, East at the right. Numbers correspond to the
identification number of the galaxies as in Table 2.
\parn
{\bf Plate 3:}
\parn
Finding chart for the galaxies with new redshift data in the OPTOPUS field
$\# 4$. Reproduced from the ESO Sky Survey
R plates; North is at the top, East at the right. Numbers correspond to the
identification number of the galaxies as in Table 2.
\parn
a) first observation; b) second observation.
\parn
{\bf Plate 4:}
\parn
Finding chart for the galaxies with new redshift data in the OPTOPUS field
$\# 5$. Reproduced from the ESO Sky Survey
R plates; North is at the top, East at the right. Numbers correspond to the
identification number of the galaxies as in Table 2.
\parn
a) first observation; b) second observation.
\parn
{\bf Plate 5:}
\parn
Finding chart for the galaxies with new redshift data in the OPTOPUS field
$\# 6$. Reproduced from the ESO Sky Survey
R plates; North is at the top, East at the right. Numbers correspond to the
identification number of the galaxies as in Table 2.
\parn
{\bf Plate 6:}
\parn
Finding chart for the galaxies with new redshift data in the OPTOPUS field
$\# 7$. Reproduced from the ESO Sky Survey
R plates; North is at the top, East at the right. Numbers correspond to the
identification number of the galaxies as in Table 2.
\parn
{\bf Plate 7:}
\parn
Finding chart for the galaxies with new redshift data in the OPTOPUS field
$\# 8$. Reproduced from the ESO Sky Survey
R plates; North is at the top, East at the right. Numbers correspond to the
identification number of the galaxies as in Table 2.
\parn
{\bf Plate 8:}
\parn
Finding chart for the galaxies with new redshift data in the OPTOPUS field
$\# 9$. Reproduced from the ESO Sky Survey
R plates; North is at the top, East at the right. Numbers correspond to the
identification number of the galaxies as in Table 2.
\parn

\vfill\eject

{\centerline {\bf REFERENCES }}
 \bigskip

\ref Allen, D.A., Norris, R.P., Staveley--Smith, L., Meadows, V.S.,
     Roche, P.F., 90, \nat 343 45.

\ref Abell, G.O., 1958, \apjsupp 3 211. [Abell catalogue]

\ref Abell, G.O., Corwin Jr., H.G., Olowin, R.P., 1989,
     \apjsupp 70 1. [ACO catalogue]

%
\ref Beers, T.C., Flynn, K., Gebhardt, K.,  1990,
    \aj 100 32.

%
\ref Day, C.S.R., Fabian, A.C., Edge, A.C., Raychaudhury, S., 1991,
  \mnras 252 394.

\ref Dressler, A., 1988, \apj 329 519.

\ref Edge, A.C., Stewart, G.C., Fabian, A.C., Arnaud, K.A., 1990,
   \mnras 245 559.

\ref Faber, S.M., Burstein, D., 1988, in \book{Large--Scale Motions
      in the Universe} proc. of the Pontifical Academic of Science Study
      Week \#27, V.C. Rubin and G.V. Coyne eds., Princeton University,
      Princeton, p.115.

\ref Gebhardt, K., Beers, T.C., 1991, \apj 383 72.

\ref Heydon--Dumbleton, N.H., Collins, C.A., MacGillivray, H.T., 1989,
      \mnras 238 379.

\ref Kurtz, M.J., Mink, D.J., Wyatt, W.F., Fabricant, D.G., Torres, G.,
     Kriss, G.A., Tonry, J.L., 1992, in \book{Astronomical Data Analysis
     Software and Systems I}, Worrall, D.M., Biemesderfer, C., and Barnes, J.,
     eds., ASP conference series vol.25, p.432.

%
\ref Lahav, O., Edge, A.C., Fabian, A.C., Putney, A., 1989,
     \mnras 238 881.

%
\ref Lynden--Bell, D., Faber, S.M., Burstein, D., Davies, R.L., Dressler, A.,
 Terlevich, R.J., Wegner, G., 1988, \apj 326 19.

\ref Lynden--Bell, D., Lahav, O., Burstein, D., 1989, \mnras 241 325.

\ref Lund, G., 1986, \book{OPTOPUS - ESO operating manual N.6}.

\ref Malumuth, E.M., Kriss G.A., Van Dyke Dixon, W., Ferguson, H.C.,
     Ritchie, C., 1992, \aj 104 495.

\ref Mathewson, D.S., Ford, V.L., Buchhorn, M., 1992,   \apjl 389 5.

\ref Melnick, J., Moles, M., 1987, {\it Rev.Mex.Astr.Astrofis.} 14,
     72.

\ref Metcalfe, N., Godwin, J.G., Spenser, S.D., 1987, \mnras 225 581.

\ref Pearson, E.S., Hartley, H.O., 1962, \book{Biometrika Tables for
Statisticians} Cambridge University press, Cambridge.

\ref Postman, M., Geller, M.J., Huchra, J.P., 1988, \aj 95 267.

\ref Raychaudhury, S., 1989, \nat 342 251.

\ref Raychaudhury, S., Fabian, A.C., Edge, A.C., Jones, C., Forman, W.,
  1991, \mnras 248 101.

\ref Scaramella, R., Baiesi--Pillastrini, G., Chincarini, G.,
 Vettolani, G., Zamorani, G.,  1989, \nat 338 562.

\ref Scaramella, R., Vettolani, G., Zamorani, G.,
1991,  \apjl 376 L1.

%
%
%
\ref Tonry, J., Davis, M., 1979, \aj  84 1511.

\ref Teague, P.F., Carter, D., Gray, P.M., 1990,
   \apjsupp 72 715.

\ref Tully, B.R., Scaramella, R., Vettolani, G., Zamorani, G., 1992,
     \apj 388 9.

\ref Vettolani, G., Chincarini, G., Scaramella, R., Zamorani, G., 1990,
   \aj 99 1709.

\ref Vettolani, G., et al., 1993, in \book{Observational Cosmology}
 G. Chicarini, A. Iovino, T. Maccacaro and D. Maccagni eds.,
 ASP Conference Series, in press.

\ref Willick, J.K., 1990,  \apjl 351 5.

\ref Wyse, R.F., Gilmore G., 1992,
	\mnras 257 1.

\ref Yentis, D.J., Cruddace, R.G., Gursky, H., Stuart, B.V., Wallin, J.F.,
     MacGillivray, H.T., Collins, C.A., 1992, in
     \book{Digitized Optical Sky Surveys} H.T. MacGillivray and E.B. Thomson
      eds., Kluwer Academic Publishers, The Netherlands, p.67.

\ref Zucca, E., Zamorani, G., Scaramella, R., Vettolani, G., 1993,
     \apj 407 470.

\vfill\eject

\bye